\documentclass{jltp}

\usepackage{graphicx} 
\usepackage{chemsym}

\title{ANDREEV REFLECTION IN \Bi_{1.8}\Pb_{0.3}\Sr_{1.9}\Ca_2\Cu_3\O_x BREAK JUNCTION}

\author{D.M. Gokhfeld$^1$, D.A. Balaev$^1$, K.A. Shaykhutdinov$^1$, \\ S.I. Popkov$^{1,2}$, M.I. Petrov$^1$}

\address{$^1$L.V. Kirensky Institute of Physics, 660036, Krasnoyarsk, Russia.\\
  $^2$M.F. Reshetnev Siberian State Aerospace University, 660014,
  Krasnoyarsk, Russia \\ E-mail: smp@iph.krasn.ru}

\begin{document}
\newcommand{\Met}{\protect\kemtkn{Met}}
\sloppy
\maketitle
\begin{abstract}
  The current-voltage characteristic (CVC) of break junction made from polycrystalline \Bi_{1.8}\Pb_{0.3}\Sr_{1.9}\Ca_2\Cu_3\O_x
   is investigated. The experimental CVC has hysteretic feature that reflects a part of a curve with negative differential resistance. The CVC is discussed within the
   framework of the K\"ummel-Gunsenheimer-Nicolsky theory considering multiple Andreev reflections
   in superconductor/normal-metal/superconductor junctions.

  PACS numbers: 75.70.Cn; 74.50.+r; 75.80.Dm

\end{abstract}
\section{INTRODUCTION}
The transport properties of polycrystalline high temperature
superconductors (HTSC) are recognized to be determined by the
intergrain boundaries which are Josephson weak links. The
investigation of CVCs should illuminate the mechanisms of current
dissipation in the networks of weak links, formed in
polycrystalline HTSC. Fundamental and applied aspects of this
topic remains actual in present\cite{1}. There are serious
problems accompanying a measurements of a CVCs of HTSC. The main
is increasing of temperature due to the selfheating effect\cite{2}
at current higher than its critical value Ic and the heat emission
on the contacts HTSC - current leads. These effects prevent a
precise measure of a CVC in a wide range of currents, including
the range $I>>I_c$ where the CVC becomes linear. Influence of the
selfheating effects can be avoided by using the break junction
technique. The formation of a microcrack in a bulk sample leads to
a significant reduction of the effective cross-section area for
charge transport. In the non-tunneling case two massive
polycrystalline banks are connected by weak links constituting
narrow part of a bulk sample, so called "bottleneck". The CVCs and
$I_c$ of the whole sample are determined by the bottleneck
cross-section area. For this reason one is possible to measure a
CVCs of break junction at bias currents that are much smaller than
$I_c$ of the bulk samples, and self-heating effects and vortex
motion should significantly reduced. Problem of preparation of the
low-resistance contacts of HTSC and current-leads is out of scope
of this article. It is under treating now and will be discussed
elsewhere. Break junctions made on unconventional superconductors
have been studied extensively (see for example
Refs.\cite{3,4,5,6,7,8,9}). The observed characteristic features
(negative differential resistance, hysteretic peculiarity,
subgarmonic gap structure, etc.) are inherent to Josephson
junctions. In this article we report on investigation of the CVC
of the break junction in polycrystalline HTSC
\Bi_{1.8}\Pb_{0.3}\Sr_{1.9}\Ca_2\Cu_3\O_x.

\section{RESULTS AND DISCUSSION}

\begin{figure}[tb]
   \centerline{\includegraphics[height=7.5cm]
   {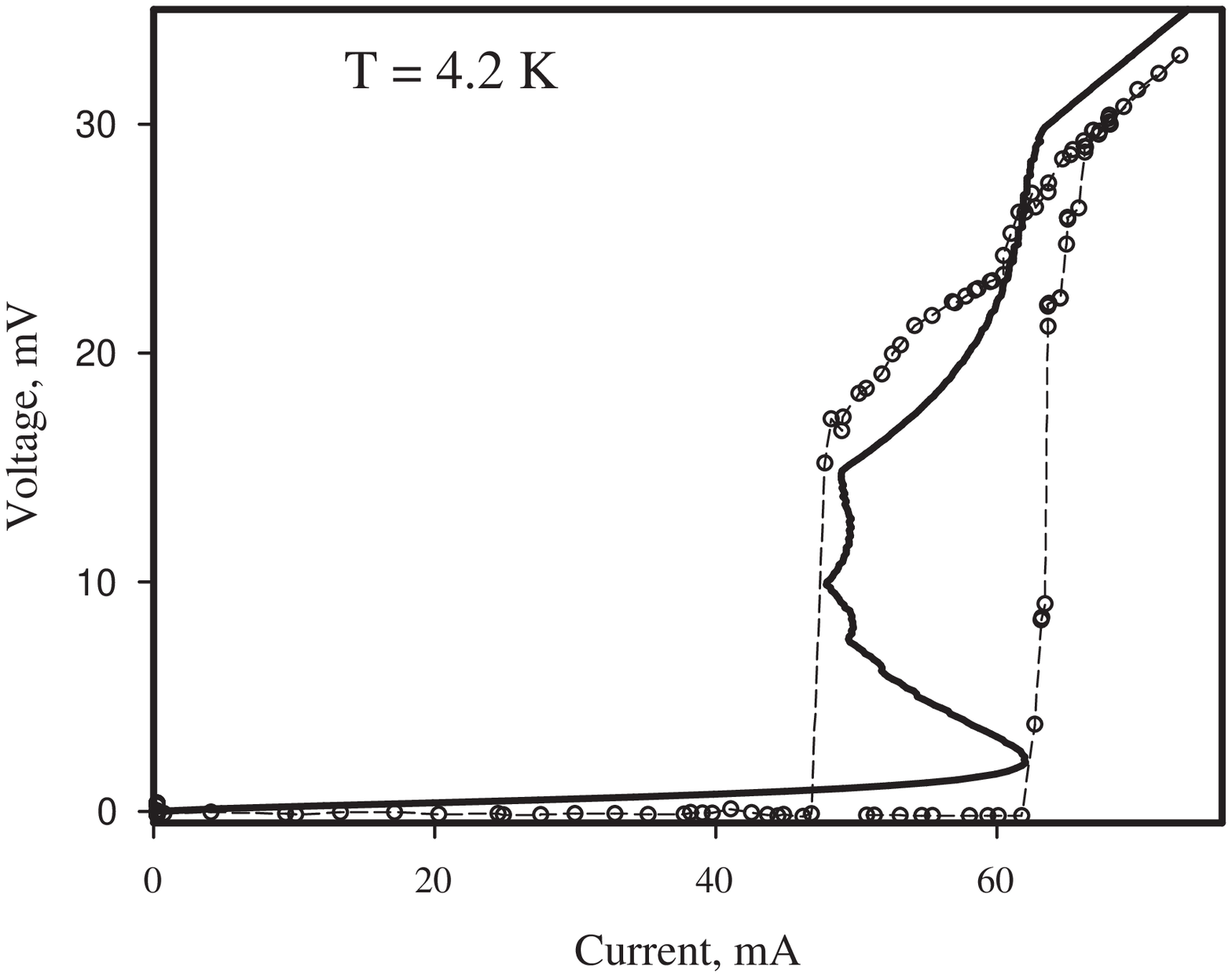}}
   \caption{Experimental (dots) and simulated (line) CVC of break junction
   \Bi_{1.8}\Pb_{0.3}\Sr_{1.9}\Ca_2\Cu_3\O_x.}
   \label{fig1}
\end{figure}

\Bi_{1.8}\Pb_{0.3}\Sr_{1.9}\Ca_2\Cu_3\O_x was prepared using the
solid state reaction technique. The preparation of the break
junction is described in Refs.\cite{7,8}. During measurement, the
sample was immersed in a liquid helium bath. The CVC measurements
were carried out under isothermal conditions by slowly scanning a
bias current. The measured CVC (Fig.~\ref{fig1}, dots) has
hysteretic peculiarity and other features, typical for
polycrystalline HTSC forming network of Josephson weak
links\cite{8,10}. It is logical to interpret the experimental data
in the frame of theories developed for dissipative transport
through S-N-S junctions (where S is superconducting bank, N is
normal region). Previously we systematically investigated the
temperature evolution of CVCs of break junctions in
polycrystalline HTSCs \Y\Ba_2\Cu_3\O_7 \cite{7} and
\La_{1.85}\Sr_{0.15}\Cu\O_4 \cite{8} and successfully described
their in the framework of the K\"ummel-Gunsenheimer-Nicolsky (KGN)
theory\cite{11}. The similar approach is used in this work. The
KGN theory\cite{11} considers current due to multiple Andreev
reflections in S-N-S junctions. According to KGN the expression
for the CVC of S-N-S junction is given by:
\begin{displaymath}
I = -\frac{1}{d} \frac{e}{m^*}\sum_k \sum_{n=1}^\infty
P_N(E_k)[f(E_k)k_e-(1-f(E_k))k_h] e^{(- n d / l)}
\end{displaymath}
\begin{equation}
\lbrace \arrowvert A_{n}^{+}(E_k+eU/2) \arrowvert^2 - \arrowvert
A_{n}^{-}(E_k-eU/2)\arrowvert^2 \rbrace + I_{Sh}
\end{equation}
where  $f(E_k)$ - Fermi distribution function; $P_N(E_k)$, from
Eq. (2.19) of Ref.\cite{11}, is the probability of finding the
quasiparticles in the N region of thickness $d$ and with inelastic
mean free path $l$ ; $e$ - charge and $m^*$ - effective mass of
electron; $n$ - number of Andreev reflections which a
quasiparticle that starts its motion from an Andreev level
characterized by the quantum-number set $k$ undergoes before it
moves out of the pair potential well; $A_{n}^{+}(E_k+eU/2)$,
$A_{n}^{-}(E_k-eU/2)$, from Eqs. (A.32), (A.33) of Ref.\cite{11} are
the probability amplitudes of n Andreev reflections of
quasiparticles with directions of propagation parallel (+) or
antiparallel (-) to the electric field; $I_{Sh} \sim U$ is the
Sharvin current, see Eq. (3.22) of Ref.\cite{11}. The simulated
curve (Fig. 1, solid line) is calculated with the following
parameters of HTSC: critical temperature $T_c$ = 112 K, energy gap
$\Delta$ = 25 meV, relation of N-region thickness to mean free
path of electrons $d/l$ = 0.09, e.g. $d$ = 10 \AA, $l$ = 110 \AA.
These parameters are physically reasonable and in good agreement
with other estimation\cite{4}.

\section{CONCLUSIONS}

Satisfactory agreement of the simulated curve and the experimental
CVC is observed. It confirms that the hysteretic peculiarity
reflects negative differential resistance range which is due to
dependence of Andreev reflections number on voltage value\cite{11}
(see also Ref.\cite{10}). As the simulated curve demonstrates, the
arch in the experimental CVC is possible the last arch of the
subgarmonic gap structure produced multiple Andreev reflections
too\cite{11}. The alike behavior was observed in the experimental
CVC of break junction in polycrystalline
\La_{1.85}\Sr_{0.15}\Cu\O_4 \cite{8}. Fig.1 demonstrates the
different curvature of the arch in the experimental CVC than one
in the calculated curve. We did not observe the such divergence in
a CVCs of polycrystalline \La_{1.85}\Sr_{0.15}\Cu\O_4 and
\Y\Ba_2\Cu_3\O_7. High anisotropy of the crystallites in
polycrystalline \Bi_{1.8}\Pb_{0.3}\Sr_{1.9}\Ca_2\Cu_3\O_x may be
responsible for it. However, instead previous
considerations\cite{7,8}, the presented experimental CVC is
described without account of a dispersion of the parameters of the
weak links in bottleneck. More surprising result is the
satisfactory agreement of the experiment on HTSCs (this article
and Refs.\cite{7,8}) and the theory\cite{11} developed for
transport through weak links based on conventional
superconductors. Further theoretical investigations involving
elastic scattering and time-dependent density functional theory
for superconductors (account of the strong electron correlations)
should be performed for more correcting comparison with
experiments in HTSC.

\section*{ACKNOWLEDGMENTS}
We are thankful to Prof. R. K\"ummel for fruitful discussions and
to Dr. V. Kravchenko for synthesis of Bi-ceramics. This work is
supported by Krasnoyarsk Regional  Scientific Foundation, grant
12F0033C.  S.I. Popkov acknowledges Krasnoyarsk Regional
Scientific Foundation for financial support, grant 14G025. K.A.
Shaykhutdinov and M.I. Petrov acknowledge program of President of
Russian Federation for support of young scientists, grant MK
1682.2004.2.

\end{document}